\documentclass[conference]{IEEEtran}
\IEEEoverridecommandlockouts
\usepackage{cite}
\usepackage{amsmath,amssymb,amsfonts}
\usepackage{algorithmic}
\usepackage{algorithm}
\usepackage{graphicx}
\usepackage{textcomp}
\usepackage{xcolor}
\usepackage{booktabs}
\usepackage{orcidlink}
\def\BibTeX{{\rm B\kern-.05em{\sc i\kern-.025em b}\kern-.08em
    T\kern-.1667em\lower.7ex\hbox{E}\kern-.125emX}}
\begin{document}

\title{
 Topology-Aware Propagation-Based Assessment of Extreme-Weather Impacts on Distribution System Resilience\\
\thanks{This paper has been accepted for presentation at the 2026 IEEE North American Power Symposium (NAPS 2026). The final version will appear in IEEE Xplore.\\ *corresponding author: Fangxing (Fran) Li (fli6@utk.edu)}
}

\author{\IEEEauthorblockN{Junjie Yin\,\orcidlink{0000-0001-7782-7274},~\IEEEmembership{Graduate Student Member,~IEEE,}
Xinyu Feng\,\orcidlink{0000-0003-4020-3219},~\IEEEmembership{Graduate Student Member,~IEEE,}\\
Jian Han\,\orcidlink{0009-0009-5612-0160}, ~\IEEEmembership{Graduate Student Member,~IEEE,}
Fangxing (Fran) Li\,{*}\,\orcidlink{0000-0003-1060-7618}~\IEEEmembership{Fellow,~IEEE}}
\IEEEauthorblockA{\textit{Department of Electrical Engineering and
Computer Science} \\
\textit{University of Tennessee}\\
Knoxville, Tennessee 37996 \\
\{jyin10, xfeng15, jhan33\}@vols.utk.edu, fli6@utk.edu }
}

\maketitle

\begin{abstract}
Extreme weather events and the increasing integration of distributed energy resources (DERs) introduce growing uncertainty and resilience challenges for distribution systems. Unlike conventional deterministic contingencies, weather-driven disruptions exhibit probabilistic and spatial-temporal characteristics, where outage consequences depend on both geographic exposure and feeder topology. Existing approaches commonly focus on deterministic outage analysis, while topology-aware operational impact assessment under forecast uncertainty remains limited. This paper proposes an event-conditioned uncertainty modeling and topology-aware impact propagation framework for distribution systems under torrential rain events. The proposed framework integrates probabilistic event-track modeling, branch-level fault screening, downstream impact propagation analysis, and operational impact assessment within a unified workflow.  Finally, case studies on the IEEE 33-bus distribution feeder demonstrate that the proposed framework can distinguish geographic exposure from topology-dependent operational impacts and support progressive early-warning impact assessment under uncertain scenarios. Furthermore, the impact zones are visualized on the CURENT Large-scale Testbed (LTB)-AGVis platform.
\end{abstract}


\section{Introduction}

Extreme weather events, including hurricanes, floods, winter storms, wildfires, and torrential rain, have become major threats to power-system resilience and operational reliability under changing climate conditions~\cite{huangResilientModernPower2024}. At the same time, the increasing penetration of distributed energy resources (DERs) further increases operational uncertainty and system complexity~\cite{nature-resilience}. Compared with transmission systems, distribution networks are typically more geographically exposed, consist of large numbers of overhead components, and operate under radial feeder structures with limited operational redundancy and more inverter-based resources (IBRs)~\cite{jyinEncodingTSG2026, fahadDataDrivenAdaptiveControl2025a,kongComparativeStudyTransmission2023a}. Consequently, distribution systems are particularly vulnerable to geographically localized but operationally widespread disruptions caused by extreme weather events.

Existing resilience assessment studies have provided important insights into extreme-event impacts on power systems through reliability analysis, fragility modeling, outage simulation, and restoration studies~\cite{wangResilienceOrientedMultiStageAdaptive2023a,jiangReliabilityAssessmentDistribution2021}. However, many existing approaches simplify extreme-event impacts using deterministic contingency assumptions such as N-1 or N-k outage models. Although such formulations are effective for conventional reliability analysis, they may not fully capture several important characteristics of extreme weather events. First, weather-driven outages are inherently probabilistic and evolve under uncertain event trajectories and localized hazard intensities. Second, extreme events exhibit strong spatial-temporal characteristics, while the operational impact on power systems depends simultaneously on geographic exposure, network topology, and component fragility~\cite{vahediResilienceAssessmentDistribution2024}. In practice, whether a component outage occurs is jointly determined by the evolving weather conditions and the physical vulnerability of the power-system infrastructure itself~\cite{shiEnhancingDistributionSystem2022}. Therefore, directly abstracting extreme events into predefined deterministic outages may overlook the event-conditioned and spatially correlated nature of distribution-system disruptions.

Considering these challenges, recent studies have investigated weather-aware outage analysis, spatial risk assessment, impact analysis~\cite{yangPVSizerOpenSourcePython2026a}, and visualization-assisted situational awareness for distribution systems~\cite{chenCollaborativeControlStrategy2023}. These efforts have significantly improved the understanding of how extreme events interact with power infrastructure. Nevertheless, from the perspective of utility early-warning operation, operators are often more concerned with identifying potential impact zones and downstream operational consequences associated with forecasted event evolution. Existing visualization-oriented approaches commonly overlay weather conditions, outage reports, or infrastructure exposure on geographic maps, but may not explicitly capture how local component failures propagate through radial feeder topology and affect downstream service regions~\cite{paulQLearningBasedImpactAssessment2020}. In addition, some existing propagation-analysis approaches discretize geographic regions into generic spatial grids, which may not fully utilize the inherent electrical topology and feeder dependency structure of distribution systems. As a result, the integration of probabilistic event forecasting, topology-aware downstream propagation assessment, and operational impact quantification for early-warning applications remains relatively limited~\cite{aliPathwayOpenSource2025, yinAIedupaper2026}.

To address these gaps, this paper proposes an event-conditioned uncertainty modeling framework and a topology-aware impact propagation framework for distribution systems under torrential rain events.
 Section~\ref{sec2} presents the probabilistic torrential rain event modeling and event-conditioned branch fault probability analysis. Section~\ref{sec3} develops the topology-aware downstream impact propagation and operational impact assessment framework under radial feeder structures. Section~\ref{sec4} demonstrates the proposed framework on the CURENT Large-scale Testbed (LTB)-AGVis platform for visualization-assisted validation and early-warning situational awareness. The main contributions of this paper are summarized as follows:
\begin{itemize}
\item An event-conditioned uncertainty modeling workflow is developed to represent torrential rain track uncertainty and spatial event influence for distribution systems.
\item A topology-aware impact propagation framework is proposed to map exposed distribution branches into downstream affected impact zones.
\item A unified operational impact assessment framework is developed to quantify both service interruption consequences and post-contingency network performance.
\item The proposed framework is validated on the IEEE 33-bus feeder and demonstrated through interactive visualization on the CURENT LTB-AGVis platform to support early-warning situational awareness.
\end{itemize}

\section{Event-Conditioned Uncertainty Modeling}\label{sec2}

\begin{figure*}[!htbp]
    \centering
    \includegraphics[width=\linewidth]{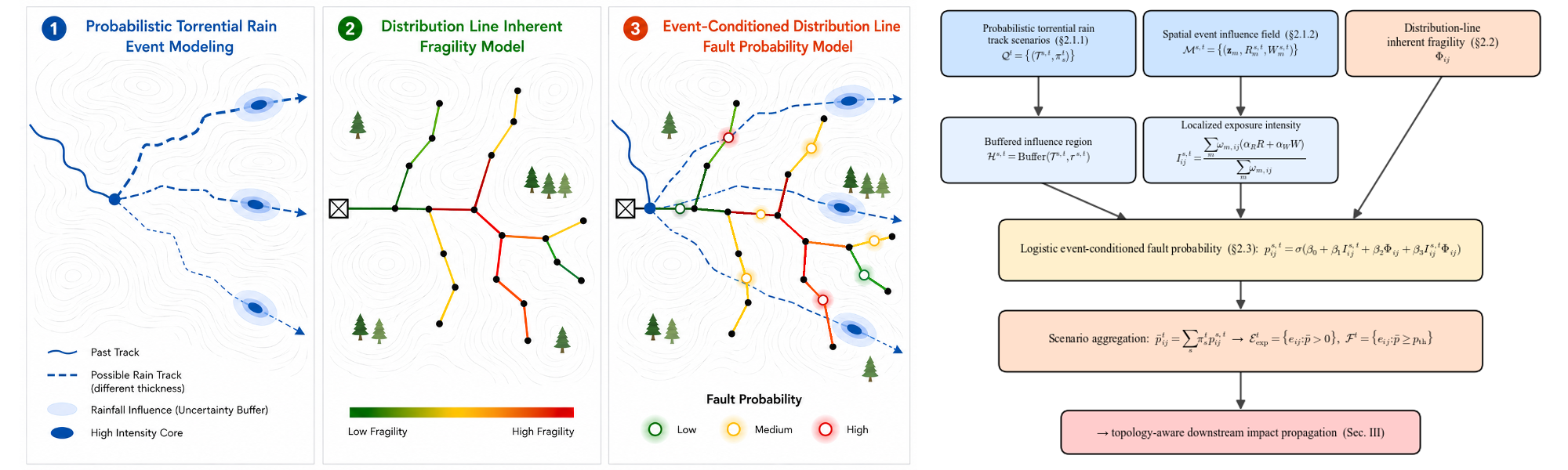}
    \caption{Workflow of the proposed event-conditioned distribution line fault probability modeling approach.}
    \label{fig:event_conditioned_fault_probability_framework}
\end{figure*}

Fig.~\ref{fig:event_conditioned_fault_probability_framework} illustrates the overall modeling workflow, in which probabilistic rain tracks and spatial influence fields characterize event uncertainty, and geography-aware fragility is incorporated to derive branch fault probabilities.

\subsection{Probabilistic Torrential Rain Event Modeling}

This paper models torrential rain as a probabilistic and spatially distributed extreme event. The proposed representation consists of two complementary components:

\subsubsection{Probabilistic Torrential Rain Track Model}

At forecast lead time $t$, the torrential rain event is represented by a set of probabilistic trajectory scenarios $\mathcal{Q}^{t}=\{(\mathcal{T}^{s,t},\pi_s^t)\}_{s\in\mathcal{S}}$, where $\mathcal{T}^{s,t}$ denotes the predicted torrential rain trajectory under scenario $s$, and $\pi_s^t$ denotes the occurrence probability associated with that scenario.

The trajectory of scenario $s$ is represented by a sequence of geographic points $\mathcal{T}^{s,t}=\{\mathbf{h}_1^{s,t},\mathbf{h}_2^{s,t},\ldots,\mathbf{h}_K^{s,t}\}$, where $\mathbf{h}_k^{s,t}$ denotes the $k$th geographic point along the predicted trajectory.

The scenario probabilities satisfy $\sum_{s\in\mathcal{S}}\pi_s^t=1$ and $\pi_s^t\geq0$.

To characterize the geographic uncertainty corridor around the predicted trajectory, a buffer-based influence region is defined as $\mathcal{H}^{s,t}=\mathrm{Buffer}(\mathcal{T}^{s,t},r^{s,t})$, where $r^{s,t}$ denotes the influence radius associated with scenario $s$ and lead time $t$. Larger values of $r^{s,t}$ correspond to longer forecast horizons with higher spatial uncertainty.

\subsubsection{Spatial Event Influence Field Model}

Since torrential rain intensity may exhibit significant spatial variability, the event is further represented as a spatial influence field over the geographic region of interest.

For scenario $s$ and lead time $t$, the spatial event influence field is defined as $\mathcal{M}^{s,t}=\{(\mathbf{z}_m,R_m^{s,t},W_m^{s,t})\}_{m=1}^{M}$, where $\mathbf{z}_m$ denotes the geographic coordinate of grid point $m$, $R_m^{s,t}$ denotes the precipitation intensity at that location, $W_m^{s,t}$ denotes the wind speed or storm-related auxiliary weather variable, and $M$ is the total number of grid points in the spatial influence field.

\subsection{Distribution Line Inherent Fragility Model}

The physical vulnerability of distribution lines is represented through an inherent fragility index, which reflects the fact that different distribution branches may exhibit different failure tendencies under the same weather exposure conditions.

The distribution system is modeled as a graph $\mathcal{G}(\mathcal{V},\mathcal{E})$, where $\mathcal{V}$ and $\mathcal{E}$ denote the bus set and branch set, respectively.

For each branch $e_{ij}\in\mathcal{E}$, the inherent fragility feature vector is defined as $\boldsymbol{\phi}_{ij}=[\phi_{ij}^{L},\phi_{ij}^{V},\phi_{ij}^{A},\phi_{ij}^{T},\phi_{ij}^{C}]$, where $\phi_{ij}^{L}$, $\phi_{ij}^{V}$, $\phi_{ij}^{A}$, $\phi_{ij}^{T}$, and $\phi_{ij}^{C}$ denote the line-length factor, vegetation exposure factor, aging factor, terrain exposure factor, and protection/criticality factor, respectively.

For standard test feeders without detailed asset-condition data, the line-length factor is approximated using normalized branch resistance $\phi_{ij}^{L}=r_{ij}/\max_{e_{mn}\in\mathcal{E}}r_{mn}$, where $r_{ij}$ is the resistance of branch $e_{ij}$.

The remaining factors are assigned normalized values within $[0,1]$, where larger values indicate higher vulnerability conditions.

The overall inherent fragility index is calculated as:
\begin{equation}
\label{eq:fragility_index}
\Phi_{ij}
=
w_L\phi_{ij}^{L}
+
w_V\phi_{ij}^{V}
+
w_A\phi_{ij}^{A}
+
w_T\phi_{ij}^{T}
+
w_C\phi_{ij}^{C},
\end{equation}
where $w_L$, $w_V$, $w_A$, $w_T$, and $w_C$ are weighting coefficients satisfying $w_L+w_V+w_A+w_T+w_C=1$.

\subsection{Event-Conditioned Distribution Line Fault Probability Model}

For each scenario $s$ and lead time $t$, a branch is first considered potentially exposed if it intersects the track-based influence corridor, i.e., $\mathcal{E}_{\mathrm{trk}}^{s,t}=\{e_{ij}\in\mathcal{E}\mid e_{ij}\cap\mathcal{H}^{s,t}\neq\emptyset\}$.

For branches within the influence corridor, localized weather exposure is further evaluated using the spatial event influence field. To capture the distance-decaying effect of localized exposure, a spatial weighting function is defined as:
\begin{equation}
\label{eq:decay}
\omega_{m,ij}
=
\exp
\left(
-\frac{
d(\mathbf{z}_m,e_{ij})
}{\lambda}
\right),
\end{equation}
where $d(\mathbf{z}_m,e_{ij})$ denotes the minimum geographic distance between grid point $\mathbf{z}_m$ and branch $e_{ij}$, and $\lambda$ is the distance-decay parameter.

The localized exposure intensity of branch $e_{ij}$ under scenario $s$ and lead time $t$ is calculated as:
\begin{equation}
\label{eq:exposure}
I_{ij}^{s,t}
=
\frac{
\sum_{m=1}^{M}
\omega_{m,ij}
\left(
\alpha_R R_m^{s,t}
+
\alpha_W W_m^{s,t}
\right)
}{
\sum_{m=1}^{M}
\omega_{m,ij}
},
\end{equation}
where $\alpha_R$ and $\alpha_W$ denote weighting coefficients associated with precipitation intensity and wind speed, respectively.

The scenario-conditioned fault probability of branch $e_{ij}$ is modeled as:
\begin{equation}
\label{eq:logistic}
p_{ij}^{s,t}
=
\sigma
\left(
\beta_0
+
\beta_1 I_{ij}^{s,t}
+
\beta_2 \Phi_{ij}
+
\beta_3 I_{ij}^{s,t}\Phi_{ij}
\right),
\end{equation}
where $\sigma(x)=1/(1+\exp(-x))$ is the logistic function; and the $\beta_{0, 1, 2, 3}$  are model coefficients
associated with the weather-exposure term, fragility term, and their interaction.

The aggregated fault probability is $\bar{p}_{ij}^{t}=\sum_{s\in\mathcal{S}}\pi_s^t p_{ij}^{s,t}$.

The exposed branch set and high-risk branch set are defined as $\mathcal{E}_{\mathrm{exp}}^{s,t}=\{e_{ij}\in\mathcal{E}\mid p_{ij}^{s,t}>0\}$ and $\mathcal{F}^{s,t}=\{e_{ij}\in\mathcal{E}\mid p_{ij}^{s,t}\geq p_{\mathrm{th}}\}$, respectively.

The resulting per-scenario fault branch set $\mathcal{F}^{s,t}$ is then passed to the downstream impact propagation module for operational impact assessment and impact-zone visualization.

\section{Topology-Aware Impact Propagation Framework}\label{sec3}
Building upon the event-conditioned fault probability model, the proposed framework performs topology-aware downstream impact propagation and operational assessment under extreme events. The framework assumes a radial feeder topology with tie-switches open, under which the level-comparison rule used in downstream propagation uniquely determines the child-side node for each faulted branch. The overall workflow is summarized in Algorithm~\ref{alg:impact_propagation}.

\begin{algorithm}[!htbp]
\caption{Topology-Aware Impact Propagation Assessment}
\label{alg:impact_propagation}
\begin{algorithmic}[1]
\REQUIRE
Power system topology,
feeder root $v_r$,
event-conditioned line fault probabilities
$\{p_{ij}^{s,t}\}$,
scenario probabilities $\{\pi_s^t\}$,
and load data $\{P_i,Q_i\}$
\ENSURE
Affected downstream bus set
$\mathcal{S}_{\mathrm{union}}^{s,t}$,
operational metrics,
and impact zone
$\mathcal{Z}_{\mathrm{imp}}^{s,t}$
\STATE Construct graph representation
$\mathcal{G}(\mathcal{V},\mathcal{E})$
\STATE Run BFS from feeder root $v_r$
to assign node levels and parent-child relationships
\FOR{each event scenario $s\in\mathcal{S}$ and lead time $t$}
\STATE Determine fault branch set
$\mathcal{F}^{s,t}$
from line fault probabilities
$\{p_{ij}^{s,t}\}$
\STATE Initialize
$\mathcal{S}_{\mathrm{union}}^{s,t}\leftarrow\emptyset$
\FOR{each fault branch
$e_k=(u_k,v_k)\in\mathcal{F}^{s,t}$}
\IF{$\ell(u_k)<\ell(v_k)$}
\STATE $c_k\leftarrow v_k$
\ELSE
\STATE $c_k\leftarrow u_k$
\ENDIF
\STATE Traverse descendants from $c_k$
to obtain downstream subtree $\mathcal{S}_k$
\STATE Update
$\mathcal{S}_{\mathrm{union}}^{s,t}
\leftarrow
\mathcal{S}_{\mathrm{union}}^{s,t}
\cup
\mathcal{S}_k$
\ENDFOR
\STATE Calculate
$P_{\mathrm{aff}}^{s,t}$,
$ENS^{s,t}$,
and
$\Gamma_V^{s,t}$
\STATE Collect coordinates of affected buses
\STATE Construct convex-hull impact region
$\mathcal{Z}_{\mathrm{imp}}^{s,t}$
\ENDFOR
\STATE Calculate expected energy not served
$EENS^t$
\RETURN
$\mathcal{S}_{\mathrm{union}}^{s,t}$,
$ENS^{s,t}$,
$EENS^t$,
$\Gamma_V^{s,t}$,
and
$\mathcal{Z}_{\mathrm{imp}}^{s,t}$
\end{algorithmic}
\end{algorithm}

\subsection{Topology-Aware Downstream Propagation}

A breadth-first-search (BFS) traversal is performed on $\mathcal{G}(\mathcal{V},\mathcal{E})$ from the feeder root to assign the node level $\ell(v_i)$ and parent-child relationships.

For each faulted branch
$e_k=(u_k,v_k)\in\mathcal{F}^{s,t}$,
the downstream child node is determined according to:
\begin{equation}
c_k=
\begin{cases}
v_k, & \ell(u_k)<\ell(v_k),\\
u_k, & \text{otherwise}.
\end{cases}
\end{equation}

The downstream affected subtree associated with branch $e_k$
is recursively extracted as:
\begin{equation}
\mathcal{S}_k
=
\{c_k\}
\cup
\bigcup_{v_j\in\mathcal{C}(c_k)}
\mathcal{S}(v_j).
\end{equation}

For multiple faulted branches,
the total affected downstream bus set is:
\begin{equation}
\mathcal{S}_{\mathrm{union}}^{s,t}
=
\bigcup_{e_k\in\mathcal{F}^{s,t}}
\mathcal{S}_k.
\end{equation}

\subsection{Fault-Induced Operational Impact Assessment}

Based on the extracted downstream affected region,
the impacted active load and energy not served (ENS)
under scenario $s$ and lead time $t$ are calculated as:
\begin{equation}
P_{\mathrm{aff}}^{s,t}
=
\sum_{i\in\mathcal{S}_{\mathrm{union}}^{s,t}}P_i,
\qquad
ENS^{s,t}
=
P_{\mathrm{aff}}^{s,t}\Delta T^{s,t}.
\end{equation}

For probabilistic early-warning assessment,
the expected energy not served (EENS) is:
\begin{equation}
EENS^t
=
\sum_{s\in\mathcal{S}}
\pi_s^t ENS^{s,t}.
\end{equation}

To evaluate post-fault operational conditions,
power flow analysis is performed on the remaining connected network.
Voltage violation severity is quantified by:
\begin{equation}
\Gamma_V^{s,t}
=
\sum_{i\in\mathcal{V}_{\mathrm{con}}^{s,t}}
\left[
\max(0,V_{\min}-V_i^{s,t})
+
\max(0,V_i^{s,t}-V_{\max})
\right],
\end{equation}
where
$\mathcal{V}_{\mathrm{con}}^{s,t}$
denotes the remaining connected bus set.

\subsection{Convex-Hull-Based Impact Zone Construction}

The topology-aware impact zone is the convex hull over the coordinates of affected buses, $\mathcal{Z}_{\mathrm{imp}}^{s,t}=\mathrm{ConvHull}(\{(x_i,y_i)\mid v_i\in\mathcal{S}_{\mathrm{union}}^{s,t}\})$.

In a radial, loop-free distribution network without DERs, each bus is supplied through a single upstream path, so a disconnection at any upstream point de-energizes the entire downstream portion of the feeder. The affected-bus set $\mathcal{S}_{\mathrm{union}}^{s,t}$ thus coincides with the de-energized downstream region, and its convex hull faithfully delineates the affected area. The BFS-based topology traversal requires $\mathcal{O}(|\mathcal{V}|+|\mathcal{E}|)$ complexity, while convex-hull construction over $N_p$ affected buses requires $\mathcal{O}(N_p\log N_p)$, ensuring computational efficiency for scenario-based early warning.

\section{Case Studies}\label{sec4}

The proposed framework is evaluated on the IEEE 33-bus distribution feeder (33 buses, 32 branches, 3.715~MW total load, maximum feeder depth 17, Bus~1 as root).

A hypothetical torrential rain event is considered to emulate a forecasted extreme-weather condition. Three possible rain tracks are generated to represent event-path uncertainty, as summarized in Table~\ref{tab:rain_scenarios}. Each scenario is associated with a probability $\pi_s$ and an influence radius $r^s$. The fault screening threshold is set as $p_{\mathrm{th}}=0.4$, and a constant outage duration of 2~h is assumed, while the framework readily supports scenario-dependent or stochastic $\Delta T^{s,t}$.

\begin{table}[!htbp]
\centering
\caption{Forecasted torrential rain scenarios}
\label{tab:rain_scenarios}
\begin{tabular}{c c c c}
\toprule
Scenario & Description & $\pi_s$ & $r^s$ \\
\midrule
A & Northern path & 0.60 & 0.50 \\
B & Central path  & 0.30 & 0.55 \\
C & Southern path & 0.10 & 0.55 \\
\bottomrule
\end{tabular}
\end{table}

The model parameters summarized in Table~\ref{tab:model_params} are configured as illustrative demonstration settings rather than field-calibrated values, and in practice they can be calibrated from historical weather and outage records using standard logistic regression, which is left for future work.

\begin{table}[!htbp]
\centering
\caption{Model parameter settings used in the case study}
\label{tab:model_params}
\resizebox{\columnwidth}{!}{%
\begin{tabular}{l l c}
\toprule
Parameter & Description & Value \\
\midrule
$[w_L,w_V,w_A,w_T,w_C]$ & Fragility weights~\eqref{eq:fragility_index} & {$[0.30,0.25,0.20,0.15,0.10]$} \\
$[\alpha_R,\alpha_W]$ & Exposure weights~\eqref{eq:exposure} & {$[0.70,0.30]$} \\
$\lambda$ & Distance-decay~\eqref{eq:decay} & {$0.30$} \\
$[\beta_0,\beta_1,\beta_2,\beta_3]$ & Logistic coeff.~\eqref{eq:logistic} & {$[-4.0,6.0,3.0,2.0]$} \\
$p_{\mathrm{th}}$ & Fault screening threshold & $0.40$ \\
$\Delta T$ & Outage duration & $2$~h \\
\bottomrule
\end{tabular}%
}
\end{table}

\subsection{Event-Conditioned Fault Scenario Analysis}

This subsection evaluates how forecasted event uncertainty is translated into branch-level fault risk and converted into probabilistic fault scenarios for subsequent impact analysis.



\begin{figure}[!htbp]
    \centering
    \includegraphics[width=1\linewidth]{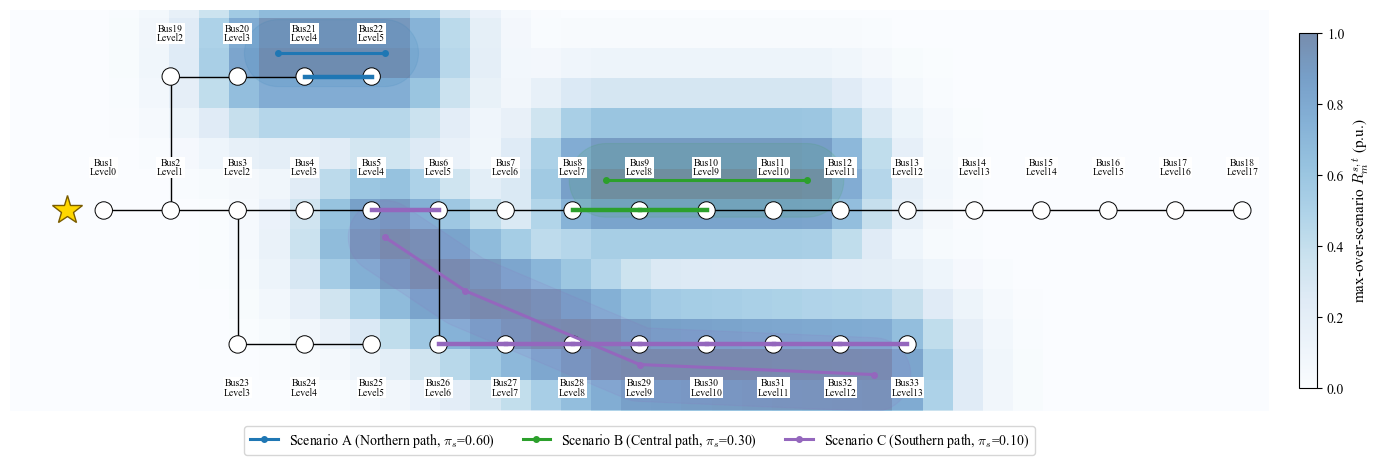}
    \caption{Aggregated event-conditioned branch fault risk on the IEEE 33-bus feeder.}
    \label{fig:aggregated_branch_risk}
\end{figure}

Combining spatial exposure with branch fragility yields the event-conditioned fault probabilities. Fig.~\ref{fig:aggregated_branch_risk} shows the aggregated branch-level fault risk over all forecasted scenarios, providing a feeder-level risk map that, unlike a deterministic hazard overlay, preserves the probabilistic contribution of multiple event paths.

After branch fault probabilities are obtained, branches with scenario-specific fault probabilities above $p_{\mathrm{th}}$ are selected as faulted branches for each scenario. Table~\ref{tab:scenario_impact} summarizes the resulting fault scenarios and their direct operational impacts. The results show that the number of geographically exposed branches does not fully determine the final system impact. For example, Scenario~C has the lowest occurrence probability but causes the largest affected load and ENS because its faulted branches are located in feeder sections with larger downstream consequences.

\begin{table}[!htbp]
\centering
\caption{Per-scenario operational impact assessment}
\label{tab:scenario_impact}
\begin{tabular}{c c c c c c}
\toprule
Scenario & $|\mathcal{E}_{\mathrm{trk}}|$ & $|\mathcal{F}^{s,t}|$ & Aff. buses & $P_{\mathrm{aff}}$ & ENS \\
 &  &  &  & (MW) & (MWh) \\
\midrule
A & 2  & 1 & 1  & 0.090 & 0.18 \\
B & 4  & 2 & 10 & 0.675 & 1.35 \\
C & 10 & 8 & 21 & 2.055 & 4.11 \\
\bottomrule
\end{tabular}
\end{table}

The resulting EENS is $\sum_s \pi_s ENS^s=0.924$~MWh.

\subsection{Topology-Aware Impact Propagation}

In a radial feeder, a faulted branch disconnects all downstream buses, so geographic exposure alone may underestimate the operational consequence of an outage.

\begin{figure}[!htbp]
    \centering
    \includegraphics[width=1\linewidth]{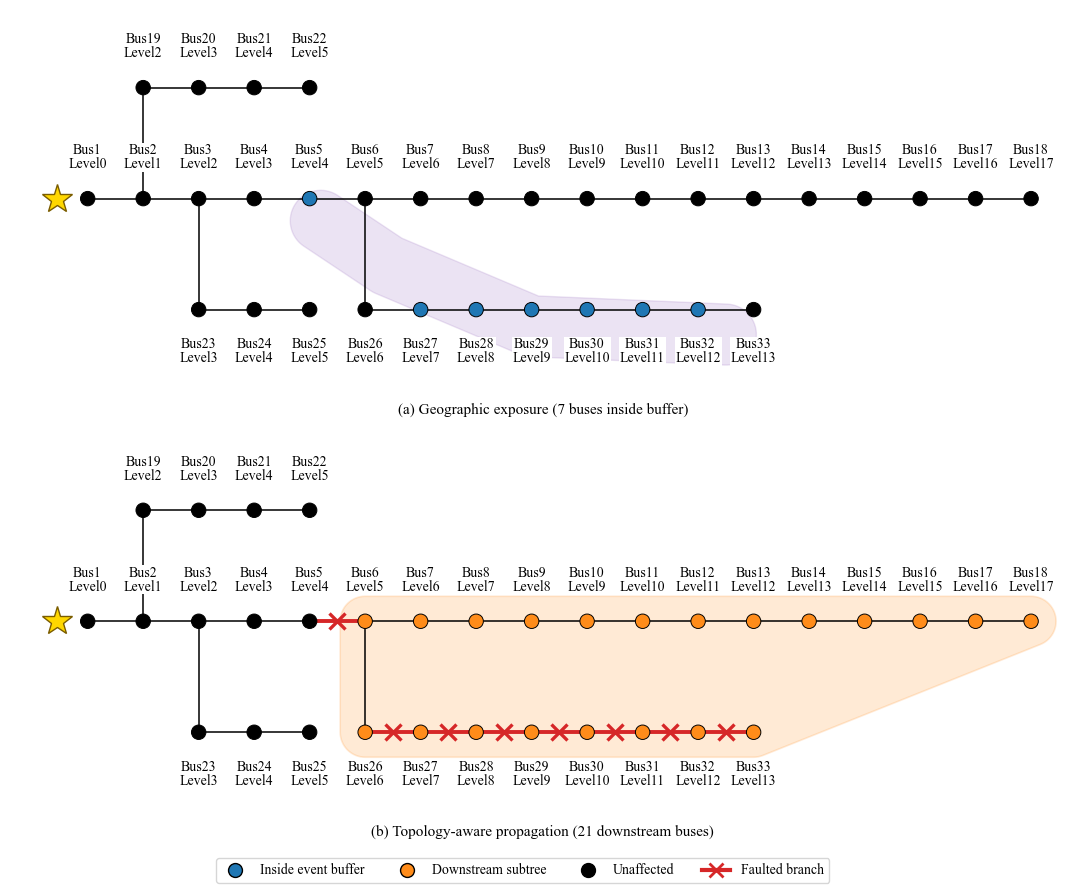}
    \caption{Comparison between geographic exposure and topology-aware downstream impact propagation.}
    \label{fig:geo_topo_comparison}
\end{figure}

\begin{figure}[!htbp]
    \centering
    \includegraphics[width=1\linewidth]{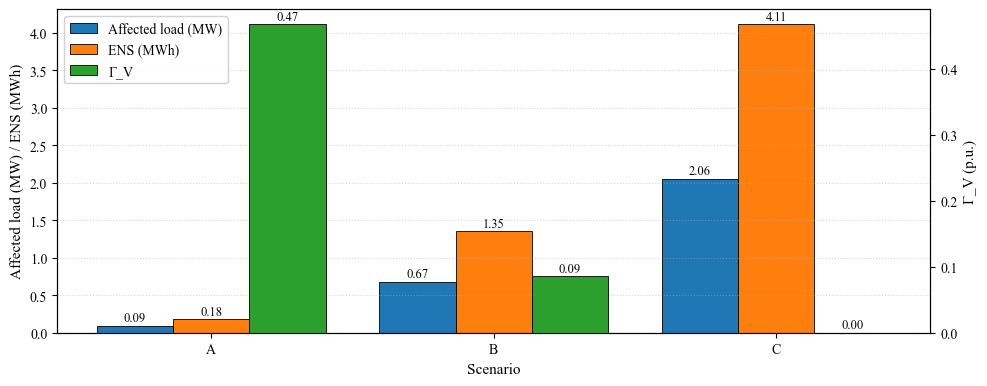}
    \caption{Scenario-wise operational impact comparison.}
    \label{fig:scenario_operational_impact}
\end{figure}

Fig.~\ref{fig:geo_topo_comparison} compares the geographically exposed region with the topology-aware downstream impact region, showing that graph-based propagation captures the electrical consequence of feeder topology rather than only spatial proximity. Fig.~\ref{fig:scenario_operational_impact} further quantifies the per-scenario impact using affected load, ENS, and voltage violation severity, where a lower-probability track (Scenario~C) can still produce the most severe consequence by intersecting upstream feeder sections, highlighting the joint role of event probability, branch fragility, and feeder topology.

\subsection{Early-Warning and Post-Impact Validation}

Since forecasted tracks and influence regions evolve with lead time, the framework aims to support progressive situational awareness rather than a single deterministic outage prediction.

\begin{figure}[!htbp]
    \centering
    \includegraphics[width=1\linewidth]{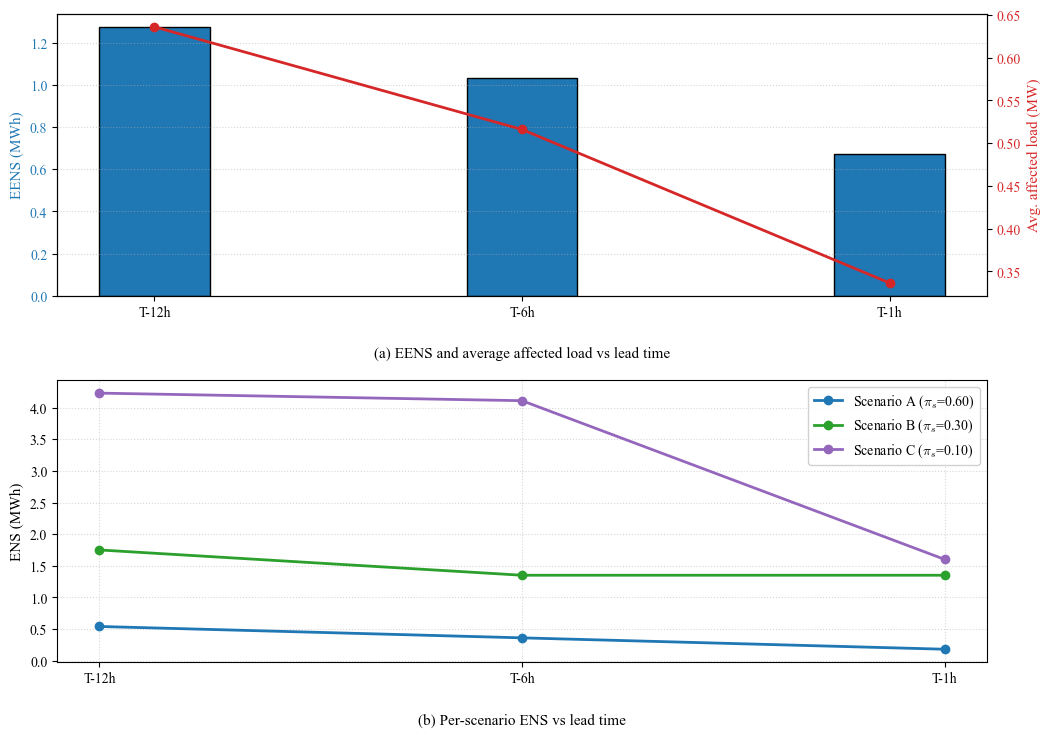}
    \caption{Time-evolving early-warning assessment under different forecast lead times.}
    \label{fig:early_warning}
\end{figure}

Fig.~\ref{fig:early_warning} summarizes the time-evolving affected load, EENS, and affected bus set under different lead times, helping operators identify whether the expected impact is increasing or decreasing as the event approaches.

\begin{figure}[!htbp]
    \centering
    \includegraphics[width=1\linewidth]{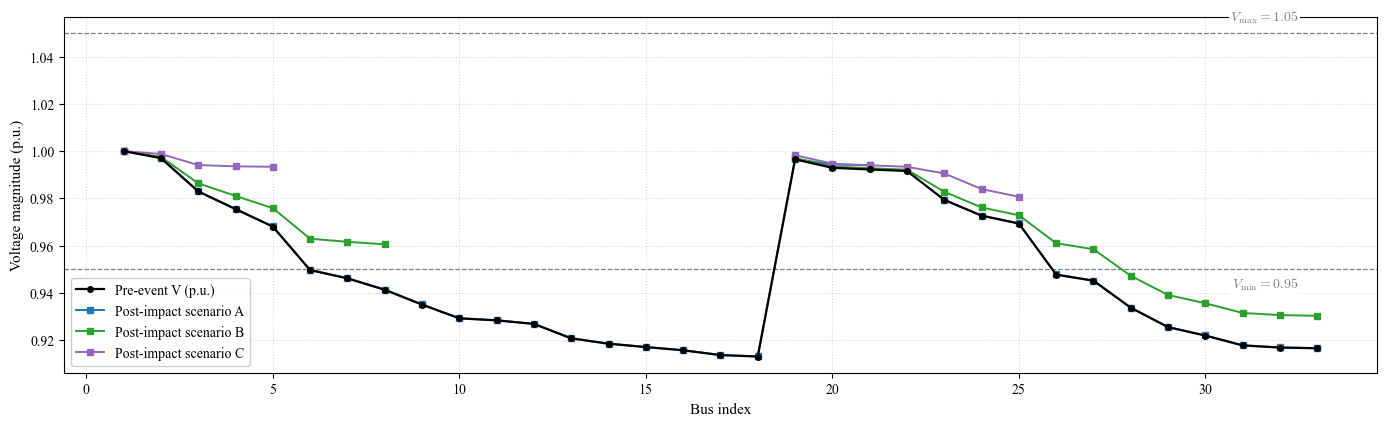}
    \caption{Pre-event and post-impact voltage profile comparison.}
    \label{fig:voltage_profile}
\end{figure}

A post-impact power flow analysis is performed after removing the faulted branches and disconnected downstream buses. Fig.~\ref{fig:voltage_profile} compares the pre-event and post-impact voltage profiles as a power-flow-based consistency check for the proposed topology-aware inference.

\subsection{CURENT LTB-AGVis Visualization of Impact Zones}

To support spatial interpretation of event tracks, exposed branches, and downstream impact zones, the inferred impact regions are exported as GeoJSON layers and rendered in CURENT LTB-AGVis through its Leaflet-based map interface, together with the identified downstream bus sets and convex-hull impact zones.

\begin{figure}[!htbp]
    \centering
    \includegraphics[width=1\linewidth]{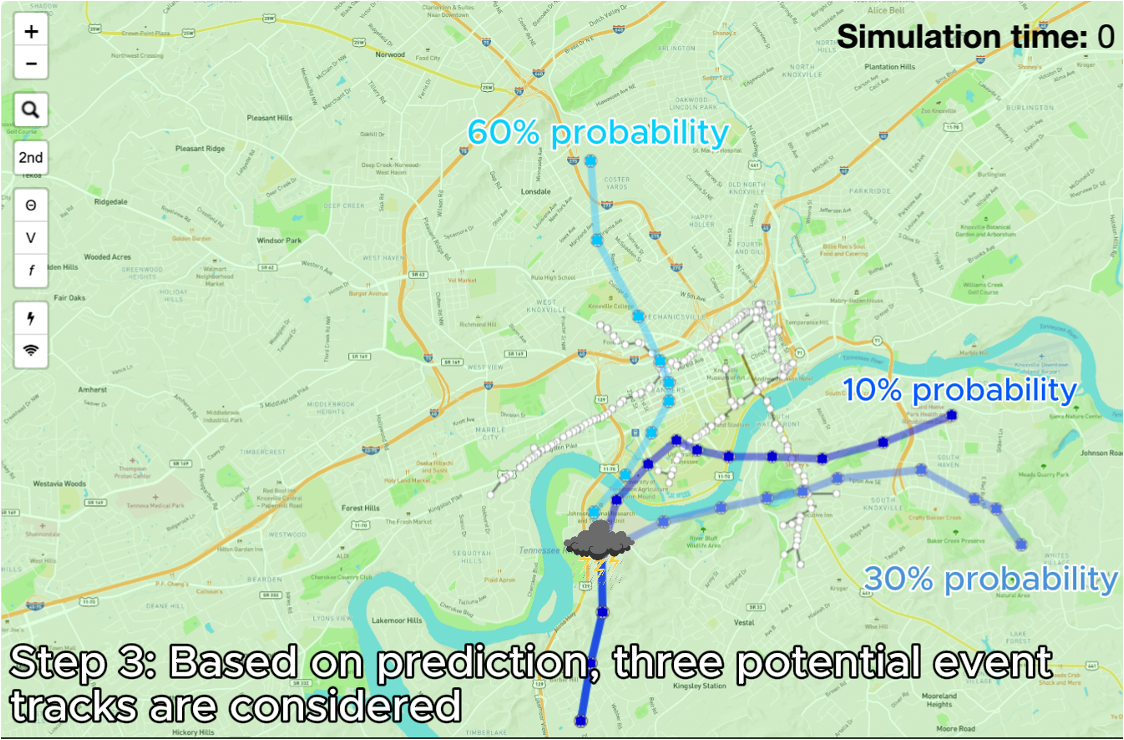}
    \caption{Potential rain track visualization in CURENT LTB-AGVis.}
    \label{fig:agvis_rain_track}
\end{figure}

Fig.~\ref{fig:agvis_rain_track} shows the forecasted rain tracks and UTK campus-based 262-bus distribution system in CURENT LTB-AGVis, allowing operators to inspect the spatial relationship between potential event paths and the feeder.

\begin{figure}[!htbp]
    \centering
    \includegraphics[width=1\linewidth]{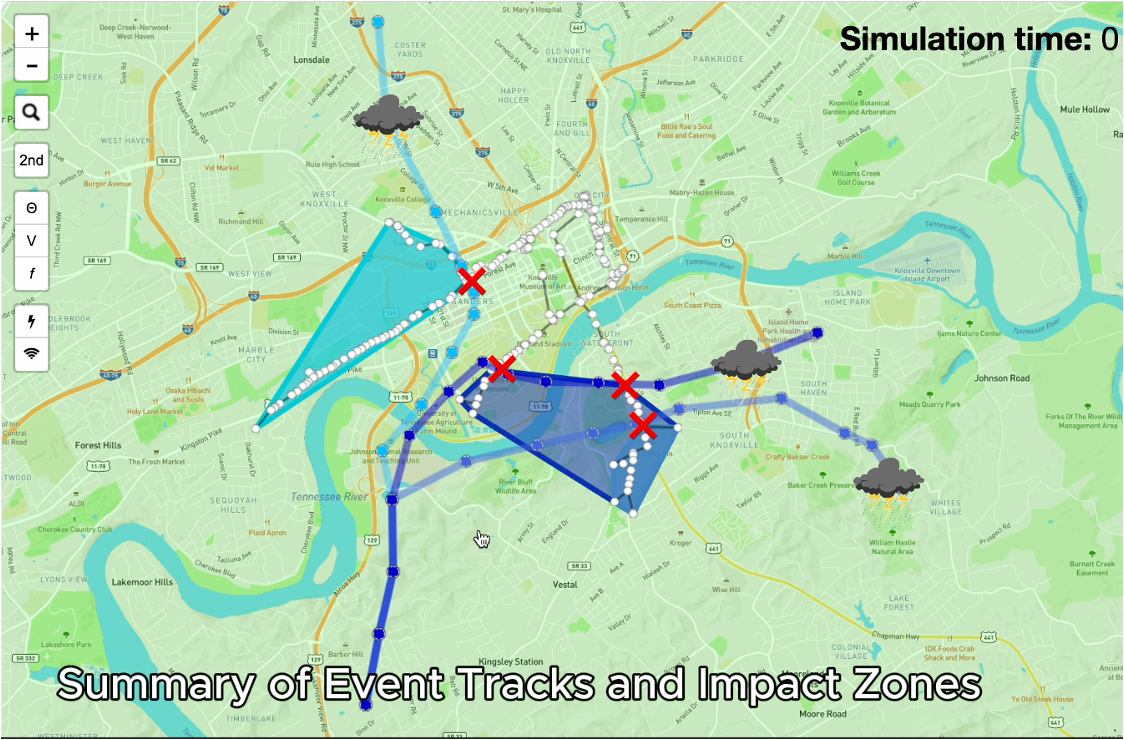}
    \caption{Impact zone validation in CURENT LTB-AGVis.}
    \label{fig:agvis_impact_zone}
\end{figure}

Fig.~\ref{fig:agvis_impact_zone} shows the topology-aware impact zones, highlighting downstream regions that may be electrically affected by faulted branches and serving as an operator-oriented validation tool for the spatial extent of potential service interruption.

For the full video, please refer to the YouTube Channel @CURENT LTB~\cite{youtube}.

\section{Conclusion and Future Work}\label{sec5}

This paper proposed an event-conditioned uncertainty modeling and topology-aware impact propagation framework for distribution systems under torrential rain events. The proposed method integrates probabilistic event-track modeling, spatial influence field construction, and topology-aware downstream impact analysis to estimate branch fault risks and quantify operational impacts under forecast uncertainty. Case studies on the IEEE 33-bus feeder demonstrated the framework's ability to support early-warning assessment under multiple scenarios. The inferred impact regions were further visualized on the CURENT LTB-AGVis platform to support operator-oriented validation and situational awareness. 

Future work will focus on integrating live weather and operational data for online impact assessment, improving extreme-event and component fragility modeling, extending the framework to meshed or looped feeder topologies, and incorporating scenario reduction methods to identify representative high-impact scenarios from large-scale simulations. Additional extensions will investigate resilience-oriented metrics, preventive control strategies, and early-warning-guided operational decision support under evolving forecast uncertainty.

\section*{Acknowledgment}

This work was supported by the CURENT Research Center.


\begin{thebibliography}{1}

\bibitem{huangResilientModernPower2024}
H.~Huang, H.~Vincent~Poor, K.~R. Davis, \textit{et al.}, ``Toward resilient modern power systems: from single-domain to cross-domain resilience enhancement,'' \emph{Proceedings of the IEEE}, vol.~112, no.~4, pp. 365--398, 2024.

\bibitem{nature-resilience}
X.~Feng, J.~Yin, C.~Li, V.~Wilson, B.~She, Q.~Zhang, J.~Zhao, C.F.~Chen, K.~Tomsovic, X.~Fang, H.~Cui, R.~Bo, F.~Li, ``Grid Resilience, Extreme Weather Events, and Clean Energy Development: Understanding Their Interactions in Modern Power Systems ,'' \emph{Nature Reviews Clean Technology}, 2026. (invited paper)

\bibitem{jyinEncodingTSG2026}
J.~Yin, B.~She, J.~Wang, F.~Li, \textit{et al.}, ``Encoding Frequency Dynamics into Economic Operation of {IBR}-Penetrated Power Systems: Quantification, Integration, and Validation,'' \emph{IEEE Transactions on Smart Grid}, 2026.

\bibitem{fahadDataDrivenAdaptiveControl2025a}
S.~Fahad, B.~She, J.~Yin, \textit{et al.}, ``A data-driven adaptive control approach for enhancing the dynamic response of {VSGs} in varying grid conditions,'' \emph{IEEE Transactions on Power Delivery}, vol.~40, no.~3, pp. 1421--1433, 2025.

\bibitem{kongComparativeStudyTransmission2023a}
B.~Kong, J.~Zhu, S.~Wang, \textit{et al.}, ``Comparative study of the transmission capacity of grid-forming converters and grid-following converters,'' \emph{Energies}, vol.~16, no.~6, pp. 2594--2594, 2023.

\bibitem{wangResilienceOrientedMultiStageAdaptive2023a}
S.~Wang and R.~Bo, ``A resilience-oriented multi-stage adaptive distribution system planning considering multiple extreme weather events,'' \emph{IEEE Transactions on Sustainable Energy}, vol.~14, no.~2, pp. 1193--1204, 2023.

\bibitem{jiangReliabilityAssessmentDistribution2021}
S.~Jiang, Y.~Wang, D.~Wang, \textit{et al.}, ``Reliability assessment of distribution network considering differentiated end-users demand for reliability,'' \emph{IOP Conference Series: Earth and Environmental Science}, vol.~645, no.~1, pp. 012026--012026, 2021.

\bibitem{vahediResilienceAssessmentDistribution2024}
S.~Vahedi, J.~Zhao, J.~Dong, \textit{et al.}, ``Resilience assessment for distribution systems during hurricanes: a learning-based framework,'' \emph{2024 IEEE Power \& Energy Society General Meeting (PESGM)}, pp. 1--5, 2024.

\bibitem{shiEnhancingDistributionSystem2022}
Q.~Shi, W.~Liu, B.~Zeng, \textit{et al.}, ``Enhancing distribution system resilience against extreme weather events: concept review, algorithm summary, and future vision,'' \emph{International Journal of Electrical Power \& Energy Systems}, vol.~138, pp. 107860--107860, 2022.

\bibitem{yangPVSizerOpenSourcePython2026a}
Y.~Yang, S.~N. Okhuegbe, J.~Zhang, \textit{et al.}, ``{PVSizer}: an open-source Python framework for {PV} and {BESS} sizing and impact analysis in distribution networks,'' \emph{IEEE Transactions on Industry Applications}, pp. 1--15, 2026.

\bibitem{chenCollaborativeControlStrategy2023}
L.~Chen, X.~Feng, C.~Wang, \textit{et al.}, ``A collaborative control strategy for {MV}-{LV} distribution networks considering multi-level access of flexible resources,'' \emph{Electric Power Systems Research}, vol.~221, pp. 109412--109412, 2023.

\bibitem{paulQLearningBasedImpactAssessment2020}
S.~Paul, F.~Ding, U.~Kumar, \textit{et al.}, ``Q-learning-based impact assessment of propagating extreme weather on distribution grids,'' \emph{2020 IEEE Power \& Energy Society General Meeting (PESGM)}, pp. 1--5, 2020.

\bibitem{aliPathwayOpenSource2025}
A.~Ali, J.~Yin, F.~Li, \textit{et al.}, ``Pathway toward an open source ecosystem in power systems: a blueprint for collaborative innovation and software sustainability,'' \emph{IEEE Electrification Magazine}, vol.~13, no.~2, pp. 58--67, 2025.

\bibitem{yinAIedupaper2026}
J.~Yin, B.~She, X.~Feng, F.~Li, ``Bridging artificial intelligence and power systems education using a hands-on executable framework,'' \emph{arXiv preprint}, 2026, \url{https://doi.org/10.48550/arXiv.2608.02599}.


\bibitem{youtube}
J.~Yin, F.~Li, ``{CURENT LTB} demo: extreme events and impact zones,'' \emph{YouTube video}, 2026, \url{https://youtu.be/Npd_iQdnOkM?si=ybWqJ7nn2a7bhNmE}.

\end{thebibliography}

\end{document}